\documentclass[prl,twocolumn,showpacs,floatfix,amsfonts]{revtex4}
\usepackage{graphicx,graphics,color,epsfig}
\usepackage{bm}
\usepackage{amsmath}
\usepackage{amssymb}
\begin{document}
\preprint{}
\title{Theory of Current and Shot Noise Spectroscopy in
Single-Molecular Quantum Dots with Phonon Mode }
\author{Jian-Xin Zhu}
\affiliation{Theoretical Division, Los Alamos National Laboratory,
Los Alamos, New Mexico 87545}
\author{A. V. Balatsky}
\affiliation{Theoretical Division, Los Alamos National Laboratory,
Los Alamos, New Mexico 87545}
\begin{abstract}
Using the Keldysh nonequilibrium Green function technique, we
study the current and shot noise spectroscopy of a single
molecular quantum dot coupled to a local phonon mode. It is found
that in the presence of electron-phonon coupling, in addition to
the resonant peak associated with the single level of the dot,
satellite peaks with the separation set by the frequency of phonon
mode appear in the differential conductance. In the ``single
level'' resonant tunneling region, the differential shot noise
power exhibit two split peaks. However, only single peaks show up
in the ``phonon assisted'' resonant-tunneling region. An
experimental setup to test these predictions is also proposed.
\end{abstract}
\pacs{73.40.Gk, 72.70.+m, 73.63.Kv, 85.65.+h} \maketitle

Recent progress in engineering and fabrication of nanoscale
electronic devices has made it possible to study the transport
properties of molecular devices~\cite{Aviram98,Langlais99,
Park00}, the linear dimension of which is at least an order
smaller than semiconductor quantum dots. In addition to their
potential industrial application, these devices provide an ideal
test ground for the study of basic physics including the quantum
size and many-body effects. The mechanism for electron conduction
on such a small scale is not well understood yet. However, the
evidence for quantum nature of transport properties has been
observed in differential conductance for molecular wires by
several experimental
groups~\cite{Andres96,Reed97,Kergueris99,Reichert02,Hong00,Rosink00,Chen00,Porath00,Smit02}.
Theoretically, a lot of effort has been focused on the study of
current-voltage characteristics of molecular wires based on either
semi-empirical~\cite{Tian98,Magoga99,Hall00,Paulsson01,Emberly00}
or {\em ab initio}~\cite{Ventra00,Taylor01,Palacios01,Damle02}
methods. In contrast to semiconductor quantum dots, which is quite
rigid in space, molecules involved in the electron tunneling
process naturally possess the vibrational degrees of freedom which
will inevitably react to the electron transfer through the
molecules. So far, the influence of inelastic scattering process
on the current-voltage characteristics of molecular wires has been
considered~\cite{Emberly00} while the influence on the current and
its fluctuation (i.e., shot noise) has not been addressed for
molecular dots. In this Letter, we use the Keldysh nonequilibrium
Green function technique to calculate the current and shot noise
through a single molecular quantum dot, for the first time.  We
focus on the effect of inelastic scattering process. A simplest
Holstein-type model with a single phonon mode is employed to
address the vibrational degrees of freedom in the molecular dot.
All other complexity of real molecular devices, apart from
interaction with phonon mode, is ignored. We find that  in
addition to the resonant peak associated with the single level of
the dot, satellite peaks with the separation set by the frequency
of phonon mode appear in the differential conductance. In the
``single level'' resonant tunneling region, the differential shot
noise power exhibits two split peaks. However, only single peaks
show up in the ``phonon assisted'' resonant-tunneling region.
Therefore, due to the quantum nature of tunneling, the Fano factor
is dramatically different from the Poisson limit even in the
presence of inelastic process. These results may also be applied
to the STM-based inelastic tunneling spectroscopy around a local
vibrational mode on surfaces~\cite{Stipe99}.

\begin{figure}
\centerline{\psfig{figure=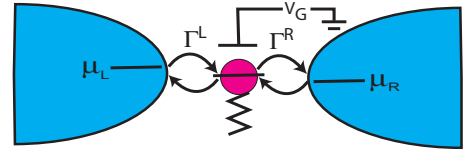,height=2cm,width=6cm}}
\caption{Schematic diagram of a molecular quantum dot system. The
dot is connected to two leads with couplings $\Gamma^{L}$ and
$\Gamma^{R}$. The dot electrons are also interacting with a single
phonon mode. A gate electrode is capacitively attached to the dot
to tune the single energy level. } \label{FIG:SETUP}
\end{figure}

The model system under consideration is illustrated in
Fig.~\ref{FIG:SETUP}. It consists of a quantum dot connected with
two normal conducting leads. The electrons on the dot is also
coupled to a single phonon mode. The single level of the dot is
tuned by a gate voltage. The system Hamiltonian is written as:
\begin{equation}
H=H_{L}+H_{R}+H_{X}+H_{D}+H_{T}\;.
\end{equation}
The first two terms are respectively the Hamiltonian for electrons
in the left and right non-interacting  metallic leads:
\begin{equation}
H_{L}+H_{R}=\sum_{k\in L,R;\sigma} \epsilon_{k}
c_{k\sigma}^{\dagger}c_{k\sigma}\;,
\end{equation}
where we have denoted the electron creation (annihilation)
operators in the leads by $c_{k\sigma}^{\dagger}$ ($c_{k\sigma}$).
The quantity $k$ is the momentum and $\sigma$ is the spin index
and  $\epsilon_{k}$ is the single particle energy of conduction
electrons. The third term describes the phonon mode:
\begin{equation}
H_{X}=\omega_{0} a^{\dagger}a\;, \end{equation} where $\omega_{0}$
is the frequency of the single phonon mode, $a^{\dagger}$ ($a$) is
phonon creation (annihilation) operator. The fourth term describes
the electron in the quantum dot:
\begin{equation} H_{D}=\sum_{\sigma}
[\epsilon_{0}+\lambda(a+a^{\dagger})]
d_{\alpha}^{\dagger}d_{\alpha}\;,
\end{equation}
where $d_{\alpha}^{\dagger}$ ($d_{\alpha}$) are the creation
(annihilation) operators of dot electrons, $\epsilon_{0}$ is the
single energy level of the dot, $\lambda$ is the coupling constant
between the dot electrons and the phonon mode. The last term
represents the coupling of the dot to the leads:
\begin{equation}
H_{T}=\sum_{k\in L,R;\sigma,\alpha} [V_{k\sigma,\alpha}
c_{k\sigma}^{\dagger}d_{\alpha} +H.c.]\;,
\end{equation}
where the tunneling matrix elements $V_{k\sigma,\alpha}$ transfer
electrons through an insulating barrier out of the dot.

The current operators from the leads to the dot are, respectively,
defined as:
\begin{equation}
\hat{I}_{L(R)}=ie\sum_{k\in
L(R);\sigma,\alpha}(V_{k\sigma,\alpha}c_{k\sigma}^{\dagger}
d_{\alpha} - H.c.)\;.
\end{equation}
The terminal current is given by $I=(\langle \hat{I}_{L}\rangle
-\langle \hat{I}_{R}\rangle)/2$, where $\langle \dots \rangle$
denotes the statistical average of physical observables. Using the
Keldysh nonequilibrium Green function
formalism~\cite{Caroli71,Keldysh65}, the terminal current can be
obtained as~\cite{Meir92,Jauho94}:
\begin{eqnarray}
I&=&\frac{ie}{4\pi} \int d\epsilon \{
\mbox{Tr}[(f_{L}\mathbf{\Gamma}^{L}-
f_{R}\mathbf{\Gamma}^{R})(\mathbf{G}^{r}
-\mathbf{G}^{a})] \nonumber \\
&&+\mbox{Tr}[ (\mathbf{\Gamma}^{L}-\mathbf{\Gamma}^{R})
\mathbf{G}^{<}]\}\;, \label{EQ:Current1}
\end{eqnarray}
where $f_{L(R)}$ are the Fermi distribution function of the left
and right leads, which has different chemical potential upon a
voltage bias $\mu_{L}-\mu_{R}=eV$, the coupling of the dot to the
leads is characterized by the parameter:
\begin{equation}
\Gamma_{\alpha\alpha^{\prime}}^{L(R)}=2\pi\sum_{\sigma}
\rho_{L(R),\sigma}(\epsilon)V_{\sigma,\alpha}^{*}(\epsilon)
V_{\sigma,\alpha^{\prime}}(\epsilon)\label{EQ:Coupling}
\end{equation}
 with
$\rho_{L(R),\sigma}$ the spin-$\sigma$ band density of states in
the two leads, $G_{\alpha\alpha^{\prime}}^{r(a)}$ and
$G_{\alpha\alpha^{\prime}}^{<}$, are the Fourier transform of the
dot electron  retarded (advanced),
$G_{\alpha\alpha^{\prime}}^{r(a)}(t,t^{\prime})=\mp i\theta(\pm t
\mp t^{\prime})\langle \{d_{\alpha}(t),
d_{\alpha^{\prime}}^{\dagger}(t^{\prime})\}\rangle$,  and lesser
Green function,
$G_{\alpha\alpha^{\prime}}^{<}(t,t^{\prime})=i\langle
d_{\alpha^{\prime}}^{\dagger}(t^{\prime})d_{\alpha}(t)\rangle$.
Note that the boldface notation indicates that the level-width
function $\mathbf{\Gamma}$ and the dot electron Green functions
are matrices in the spin space of dot electrons.

The electrical current through a device is fluctuating in time
even under a dc bias~\cite{Blanter00}. Its fluctuation is
characterized by the spectral density, which is given by the
Fourier transform $S(\omega)$ of the current correlation
function~\cite{Note1}:
\begin{equation}
S(t,t^{\prime})=\frac{1}{2}\langle \{ \Delta \hat{I}_{L}(t),
\Delta \hat{I}_{L}(t^{\prime}) \}\rangle\;,
\end{equation}
where $\Delta \hat{I}_{L}(t) = \hat{I}_{L}(t)-\langle
\hat{I}_{L}(t)\rangle$. The quantum statistical (non-equilibrium)
average involved in the current correlation can be evaluated
conveniently in the Keldysh Green function formalism. Following a
tedious but standard procedure, we obtain the spectral density of
shot noise in the zero-frequency limit:
\begin{eqnarray}
&S(\omega\rightarrow 0)= \frac{e^{2}}{2\pi} \int d\epsilon \{
-f_{L}(1-f_{L})(\mbox{Tr}[(\mathbf{\Gamma}^{L}\mathbf{G}^{r})^{2}]
&\nonumber \\
& +\mbox{Tr}[(\mathbf{\Gamma}^{L}\mathbf{G}^{a})^{2}])
+if_{L}\mbox{Tr}[\mathbf{\Gamma}^{L}\mathbf{G}^{>}]
 -i(1-f_{L})\mbox{Tr}[\mathbf{\Gamma}^{L}\mathbf{G}^{<}]
&\nonumber \\
& +f_{L}\mbox{Tr}[\mathbf{\Gamma}^{L}\mathbf{G}^{>}
\mathbf{\Gamma}^{L}(\mathbf{G}^{r}-\mathbf{G}^{a})]
+\mbox{Tr}[\mathbf{\Gamma}^{L}\mathbf{G}^{>} \mathbf{\Gamma}^{L}
\mathbf{G}^{<}] & \nonumber \\
&-(1-f_{L})
\mbox{Tr}[\mathbf{\Gamma}^{L}(\mathbf{G}^{r}-\mathbf{G}^{a})
\mathbf{\Gamma}^{L}\mathbf{G}^{<}]\}\;,& \label{EQ:Shot1}
\end{eqnarray} where $G_{\alpha\alpha^{\prime}}^{>}$
is the Fourier transform of the greater Green function,
$G_{\alpha\alpha^{\prime}}^{>}(t,t^{\prime}) =-i\langle
d_{\alpha}(t)d_{\alpha^{\prime}}^{\dagger}(t^{\prime})\rangle$.
Equation~(\ref{EQ:Shot1}) is a basic formula for the low-frequency
shot noise through a quantum dot, which can take into account the
many-body effects conveniently. It expresses the fluctuations of
current through the quantum dot, an interacting region, in terms
of the distribution functions in the leads and local properties of
the quantum dot, such as the occupation and density of states.
This formalism can be viewed as a
generalized version of the two-terminal 
shot-noise formula for the non-interacting
case~\cite{Chen91,Buttiker92}, which will become clear in the
following discussion. It can also be used to study the current
fluctuation in the case of spin-dependent transport, which is of
much interest in the study of spintronics and single spin
detection.

Once the dot electron Green functions are known, the electrical
current and shot noise can be calculated using
Eqs.~(\ref{EQ:Current1}) and (\ref{EQ:Shot1}). In the following,
we calculate these Green functions, which should be carried out in
the presence of the leads.
By performing a canonical transformation,
$\bar{H}=e^{\mathcal{S}}He^{-\mathcal{S}}$ with
$\mathcal{S}=(\lambda/\omega_0)\sum_{\alpha} d_{\alpha}^{\dagger}
d_{\alpha} (a^{\dagger}-a)$~\cite{Mahan00}, one can obtain
$\bar{H}=\bar{H}_{el}+\bar{H}_{ph}$. Here
$\bar{H}_{el}=\bar{H}_{L}+\bar{H}_{R}+\bar{H}_{D}+\bar{H}_{T}$ and
$\bar{H}_{ph}=\bar{H}_{X}$, where $\bar{H}_{L(R)}=H_{L(R)}$,
$\bar{H}_{D}=\sum_{\alpha}(\epsilon_0-\lambda^{2}/\omega_0)
d_{\alpha}^{\dagger}d_ {\alpha}$, $\bar{H}_{T}\approx H_{T}$, and
$\bar{H}_{X}=H_{X}$. Consequently, the original dot-electron Green
function can be decoupled as:
\begin{eqnarray}
G_{\alpha\alpha^{\prime}}^{r(a)}(t)&=& \mp i\theta(\pm t) \langle
\{\tilde{d}_{\alpha}(t),
\tilde{d}_{\alpha^{\prime}}^{\dagger}(0)\}\rangle_{el} \langle
X(t) X^{\dagger}(0)\rangle_{ph} \nonumber \\ &=&
\tilde{G}^{r(a)}_{\alpha\alpha^{\prime}}(t)  \langle X(t)
X^{\dagger}(0)\rangle_{ph} \;, \label{EQ:Green_ra}
\end{eqnarray}
where $\tilde{d}_{\alpha}(t)=e^{i\bar{H}_{el}t}d_{\alpha}
e^{-i\bar{H}_{el}t}$, and
$X(t)=e^{i\bar{H}_{ph}t}Xe^{-i\bar{H}_{ph}t}$ with
$X=\exp[-(\lambda/\omega_{0})(a^{\dagger}-a)]$. The
renormalization factor due to the electron-phonon interaction is
evaluated to be~\cite{Mahan00}: $\langle X(t)
X^{\dagger}(0)\rangle_{ph} =e^{-\Phi(t)}$, where
$\Phi(t)=(\lambda/\omega_0)^{2}[N_{ph}(1-e^{i\omega_0 t})
+(N_{ph}+1)(1-e^{-i\omega_0 t})]$ with $N_{ph}=[\exp(\beta
\omega_0)-1]^{-1}$. We then apply the equation-of-motion approach
to compute the Green function
$\tilde{G}_{\alpha\alpha^{\prime}}^{r(a)}$. It consists of
differentiating the Green function with respect to time, thereby
generating higher-order Green functions which eventually  have to
close (by truncation in the presence of electron-electron
interaction) the equation for the Green function
$\tilde{G}^{r(a)}$. A little algebra gives rise to the Fourier
transform:
\begin{equation}
\tilde{G}_{\alpha\alpha^{\prime}}^{r(a)}(\omega)
=\frac{\delta_{\alpha\alpha^{\prime}}}{\omega-(\epsilon_0-\Delta)
-\Sigma_{\alpha\alpha}^{r(a)}}\;,
\end{equation}
where the energy shift $\Delta=\lambda^{2}/\omega_0$ and the
retarded (advanced) self-energy due to the tunneling into the
electrical leads are given by:
\begin{equation}
\Sigma^{r(a)}_{\alpha\alpha^{\prime}}(\omega)=\sum_{k\in
L,R}\frac{\vert
V_{k\alpha,\alpha}\vert^{2}\delta_{\alpha\alpha^{\prime}}}{\omega
-\epsilon_{k}\pm i 0^{+}}\;. \label{EQ:Self-Energy}
\end{equation}
The full width of the resonance is then just the sum of elastic
couplings to the two leads which is given by
Eq.~(\ref{EQ:Coupling}). The Fourier transform of the full Green
function given by Eq.~(\ref{EQ:Green_ra}) can be obtained as:
\begin{eqnarray}
G_{\alpha\alpha^{\prime}}^{r(a)}&=&e^{-g(2N_{ph}+1)}
\sum_{l=-\infty}^{\infty}
J_{l}(2g\sqrt{N_{ph}(N_{ph}+1)})\nonumber \\
&&\times \frac{\delta_{\alpha\alpha^{\prime}}e^{l\omega_0\beta/2}}
{\omega-(\epsilon_0-\Delta)-l\omega_0
-\Sigma_{\alpha\alpha}^{r(a)}}\;, \label{EQ:GreenF} \end{eqnarray}
where $J_l(x)$ are the Bessel functions of complex argument, the
parameter $g=(\lambda/\omega_0)^{2}$. The Green function
$G_{\alpha\alpha^{\prime}}^{<(>)}$ cannot be obtained by directly
using the above equation-of-motion approach. By instead applying
the operational rules as given by Langreth~\cite{Langreth76} to
the Dyson equation for the contour-ordered Green function, one can
show the following Keldysh equation for the lesser and greater
functions~\cite{Jauho94}:
$\mathbf{G}^{<(>)}(\omega)=\mathbf{G}^{r}(\omega)
\mathbf{\Sigma}^{<(>)}_{T}(\omega)\mathbf{G}^{a}(\omega)\;.$
In the weak electron-phonon coupling limit as we are considering
here, the contribution to the self-energy
$\mathbf{\Sigma}_{T}^{<(>)}$ from the electron-phonon interaction
is negligible such that $\mathbf{\Sigma}_{T}^{<(>)}$ can be
approximated by the lesser (greater) self-energy  due to the
tunneling into the two leads:
$\mathbf{\Sigma}^{<}=i[f_{L}\mathbf{\Gamma}^{L}
+ f_{R}\mathbf{\Gamma}^{R}]\;,$ and $ \mathbf{\Sigma}^{>} =
-i[(1-f_{L})\mathbf{\Gamma}^{L} +(1-f_{R})\mathbf{\Gamma}^{R}]\;.$
This relation
enables us to rewrite the current and shot noise formulae as:
\begin{equation}
I=\frac{e}{2\pi}\int d\epsilon [f_{L}-f_{R}]
\mbox{Tr}[\mathbf{T}]\;, \label{EQ:Current2}
\end{equation}
and
\begin{eqnarray}
S&=&\frac{e^{2}}{2\pi} \int d\epsilon \{ [f_{L}(1-f_{L})
+f_{R}(1-f_{R})]\mbox{Tr}[\mathbf{T}] \nonumber \\
&& +(f_{L}-f_{R})^{2} \mbox{Tr}[(1-\mathbf{T})\mathbf{T}]\}\;,
\label{EQ:Shot2}
\end{eqnarray}
where we have defined the transmission coefficient matrix
$\mathbf{T}=\mathbf{G}^{a}\mathbf{\Gamma}^{L} \mathbf{G}^{r}
\mathbf{\Gamma}^{R}$. We remark that for the noninteracting case
(no electron-phonon and electron-electron interactions), the above
two expressions are exact. These are just the
Landauer-B\"{u}ttiker formalisms developed for the noninteracting
electron transport based on the scattering matrix theory. The
connection between the two formalisms for the current was first
established by Meir and Wingreen~\cite{Meir92}.

\begin{figure}
\centerline{\psfig{figure=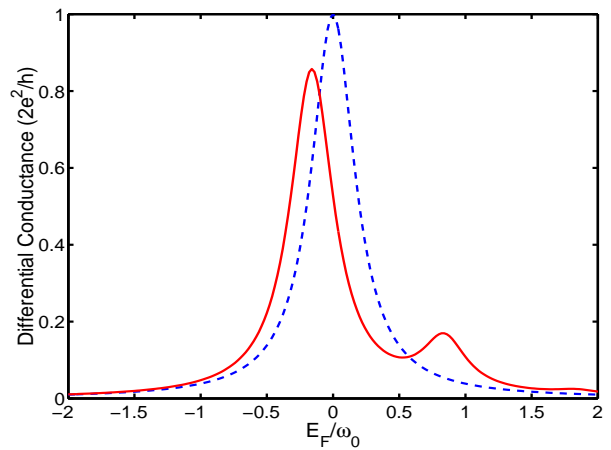,height=6cm,width=8cm}}
\caption{Differential conductance, in units of $2e^{2}/h$, through
the molecular quantum dot as a function of the Fermi energy
$E_{F}$ measured relative to the single level of the quantum dot
in the presence (red-solid line, $\lambda=0.4\omega_0$) and
absence (blue-dashed line, $\lambda=0$) of electron-phonon
coupling. The energy is measured in units of the frequency of the
phonon mode $\omega_0$, and $\Gamma^{L}=\Gamma^{R}=0.2\omega_0$. }
\label{FIG:COND}
\end{figure}

\begin{figure}
\centerline{\psfig{figure=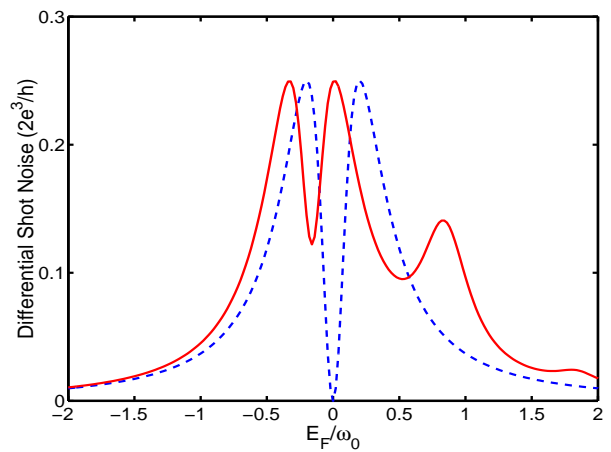,height=6cm,width=8cm}}
\caption{ Differential shot noise power, in units of $2e^{3}/h$,
through the molecular quantum dot as a function of the Fermi
energy $E_{F}$ measured relative to the single level of the
quantum dot in the presence (red-solid line,
$\lambda=0.4\omega_0$) and absence (blue-dashed line, $\lambda=0$)
of electron-phonon coupling. The energy is measured in units of
the frequency of the phonon mode $\omega_0$, and
$\Gamma^{L}=\Gamma^{R}=0.2\omega_0$. } \label{FIG:NOISE}
\end{figure}

For simplicity, we consider the coupling of the dot to the two
leads to be symmetric $\mathbf{\Gamma^{L}}=\mathbf{\Gamma}^{R}$,
and assume that the leads have broad and flat density of states
(i.e., the wideband limit). For this case, the elastic couplings,
from which the self-energy~(\ref{EQ:Self-Energy}) can be
determined, are independent of energy. In Figs.~\ref{FIG:COND} and
\ref{FIG:NOISE}, we plot the zero-temperature differential
conductance and shot noise power as a function of the Fermi energy
(that is, the equilibrium chemical potential in the leads)
measured relative to the single level $\epsilon_0$ in the presence
(with typical value $\lambda=0.4\omega_{0}$, red-solid lines in
the figures) and absence ($\lambda=0$, blue-dashed lines) of
electron-phonon coupling. Here all energy quantities are measured
in units of the phonon mode frequency $\omega_0$, and the elastic
couplings $\Gamma^{L}=\Gamma^{R}$ are fixed at $0.2\omega_0$. When
the electron is not coupled to the phonon mode, only one resonant
conductance peak shows up as the Fermi energy matches the single
level in the quantum dot. In the presence of electron-phonon
coupling, the overall spectrum is shifted by
$\Delta=\lambda^{2}/\omega_0$. In addition to the main peak
related to the single level, new satellite resonant peaks appear
at the positive energy side. The separation between the
conductance peaks is set by the frequency of the phonon mode
$\omega_0$. Since at zero temperature no phonon modes are excited
on the quantum dot, the electrons tunneling onto the dot can only
excite phonon modes, which explains why the satellite peaks are
located at the positive energy region. It can be expected that
there will appear satellite peaks at negative energy region at
finite temperatures, where thermally excited phonons can be
absorbed by the electrons tunneling onto the dot. Moreover, the
intensity of the satellite peaks is much smaller than the main
resonant peak because they are evolving from the emission of
phonon modes, which is controlled by the electron-phonon coupling.
Correspondingly, the differential shot noise power exhibits two
peaks located symmetrically around the position where the
conductance peak associated with the single level of the dot is
located. The origin of this behavior is that no noise is generated
when the transmission probability $T=0$ or $1$, while it is
generated maximally in between. In this ``single level''
resonant-tunneling region, the Fano factor defined as the ratio of
shot noise to current is significantly enhanced. Since the
transmission probability corresponding to those satellite
conductance peaks (from the phonon emission process) is small, the
differential noise power is approximately proportional to the
conductance and only show a single peak rather than two peaks.
Therefore, in the ``phonon assisted'' resonant-tunneling region,
the Fano factor is smaller than one, in contrast to the case of
``single level'' resonant tunneling. Therefore, due to the quantum
tunneling nature, the Fano factor for the present system differs
the Poisson limit (equal to 2), where electrons diffuse in an
uncorrelated way.

To experimentally investigate the current and shot noise
spectroscopy of molecular quantum dots, we suggest a setup where a
gate electrode is attached to the dot capacitively, as shown in
Fig.~\ref{FIG:SETUP}. A gate voltage is applied solely to tune the
level position on the dot while a bias voltage is applied in the
linear response regime.  The merit of the proposed setup is that
the internal electronic and ionic charge buildup on the dot, which
is generated by a nonlinear response, could be avoided safely.
Experimental effort along this direction is in
progress~\cite{Park02}.

To summarize, using the Keldysh nonequilibrium Green function
technique, we have studied in this work, to the best of our
knowledge, for the first time the current and shot noise
spectroscopy of a single molecular quantum dot coupled to a local
phonon mode. We show that, in the presence of electron-phonon
coupling, in addition to the resonant peak associated with the
single level of the dot, satellite peaks with the separation set
by the frequency of phonon mode appear in the differential
conductance. In the ``single level'' resonant tunneling region,
the differential shot noise power exhibits two split peaks.
However, only single peaks show up in the ``phonon assisted''
resonant-tunneling region. The Fano factor is different from the
Poisson limit even in the presence of inelastic process. The
derived current and noise formalism can also be applied to more
complicated systems.

{\bf Acknowledgments}: We thank A. Abanov and R. Lu for useful
discussions.  This work was supported by the Department of Energy.


\begin{thebibliography}{99}

\bibitem{Aviram98} A. Aviram and M. Ratner, eds., {\em Molecular
Electronics: Science and Technology} (Annals of the New York
Academy of Sciences, New York, 1998).

\bibitem{Langlais99} V. Langlais {\em et al.}, Phys. Rev. Lett.
{\bf 83}, 2809 (1999).

\bibitem{Park00} H. Park {\em et al.}, Nature {\bf 407}, 57 (2000).

\bibitem{Andres96} R. P. Andres {\em et al.}, Science {\bf 272},
1323 (1996).

\bibitem{Reed97} M. A. Reed {\em et al.}, Science {\bf 278}, 252
(1997).

\bibitem{Kergueris99} C. Kergueris {\em et al.}, Phys. Rev. B {\bf
59}, 12505 (1999).

\bibitem{Reichert02} J. Reichert {\em et al.}, Phys. Rev. Lett. {\bf
88}, 176804 (2002).

\bibitem{Hong00} S. Hong {\em et al.}, Superlatt. and Microstruc.
{\bf 28}, 289 (2000).

\bibitem{Rosink00} J. J. W. Rosink {\em et al.}, Phys. Rev. B {\bf
62}, 10459 (2000).

\bibitem{Chen00} J. Chen {\em et al.}, Appl. Phys. Lett. {\bf 77},
1224 (2000).

\bibitem{Porath00} D. Porath {\em et al.}, Nature {\bf 403}, 635
(2000).

\bibitem{Smit02} R. H. M. Smit {\em et al.}, cond-mat/0208407.

\bibitem{Tian98} W. Tian {\em et al.}, J. Chem. Phys. {\bf 109},
2874 (1998).

\bibitem{Magoga99} M. Magoga and C. Joachim, Phys. Rev. B {\bf
59}, 16011 (1999).

\bibitem{Hall00} L. E. Hall {\em et al.}, J. Chem. Phys. {\bf
112}, 1510 (2000).

\bibitem{Paulsson01} M. Paulsson and S. Stafstr\"{o}m, Phys. Rev.
B {\bf 64}, 035416 (2001).

\bibitem{Emberly00} E. G. Emberly and G. Kirczenow, Phys. Rev. B
{\bf 62}, 10451 (2000).

\bibitem{Ventra00} M. Di Ventra, S. T. Pantelides, and N. D. Lang,
Phys. Rev. Lett. {\bf 84}, 979 (2000).

\bibitem{Taylor01} J. Taylor, H. Gou, and J. Wang, Phys. Rev. B
{\bf 63}, 245407 (2001).

\bibitem{Palacios01} J. J. Palacios {\em et al.}, Phys. Rev. B
{\bf 64}, 115411 (2001).

\bibitem{Damle02} P. S. Damle, A. W. Ghosh, and S. Datta, Chem.
Phys. {\bf 281}, 171 (2002).

\bibitem{Stipe99} B. C. Stipe, M. A. Rezaei, and W. Ho, Science
{\bf 280}, 1732 (1998); Phys. Rev. Lett. {\bf 82}, 1724 (1999); N.
Lorente {\em et al.}, {\em ibid.} {\bf 86}, 2593 (2001).

\bibitem{Caroli71} C. Caroli {\em et al.}, J. Phys. C {\bf 4}, 916
(1971).

\bibitem{Keldysh65} L. V. Keldysh, Zh. Eksp. Teor. Fiz. {\bf 47},
1515 (1965) [Sov. Phys. JETP {\bf 20}, 1018 (1965).

\bibitem{Meir92} Y. Meir and N. S. Wingreen, Phys. Rev. Lett. {\bf
68}, 2512 (1992).

\bibitem{Jauho94} A.-P. Jauho, N. S. Wingreen, and Y. Meir, Phys.
Rev. B {\bf 50}, 5528 (1994).

\bibitem{Blanter00} For a review, see, Ya. M. Blanter and M.
B\"{u}ttiker, Phys. Rep. {\bf 336}, 1 (2000).

\bibitem{Note1} For the case of a quantum dot connected to two
leads, the current conservation requires that $\langle
\hat{I}_{L}\rangle  =-\langle \hat{I}_{R}\rangle$. Therefore, the
current fluctuations are the same at either lead, and they have
the opposite sign with the correlation of fluctuations at two
leads, i.e., $\langle \Delta \hat{I}_{L}\Delta \hat{I}_{L}\rangle
(\omega)  =\langle \Delta \hat{I}_{R}\Delta \hat{I}_{R}\rangle
(\omega)=-\langle \Delta \hat{I}_{L}\Delta \hat{I}_{R}\rangle
(\omega)$. It is then sufficient for us to calculate the current
correlation at the left lead.

\bibitem{Chen91} L. Y. Chen and C. S. Ting, Phys. Rev. B {\bf 43},
4534 (1991).

\bibitem{Buttiker92} M. B\"{u}ttiker, Phys. Rev. B {\bf 46}, 12485
(1992); Phys. Rev. Lett. {\bf 65}, 2901 (1990).

\bibitem{Mahan00} G. D. Mahan, {\em Many-Particle Physics}, 3rd
ed. (Plenum Press, New York, 2000).

\bibitem{Langreth76} D. C. Langreth, in {\em Linear and Nonlinear
Electron Transport in Solids}, edited by J. T. Devreese and V. E.
Van Doren (Plenum, New York, 1976).

\bibitem{Park02} J. Park {\em et al.}, {\em ibid.} {\bf 417},
722 (2002); W. Liang {\em et al.}, {\em ibid.} {\bf 417}, 725
(2002).

\end{thebibliography}
\end{document}